\documentclass[preprint2]{aastex}

\newcommand{\Msun}{\hbox{$\hbox{M}_\odot\;$}}
\newcommand{\Rsun}{\hbox{$\hbox{R}_\odot\;$}}
\newcommand{\kms}{\hbox{${\rm km}\:{\rm s}^{-1}\;$}}

\slugcomment{To appear in ApJ Letters}

\shorttitle{Fast orbital decay in XTE J1118+480}
\shortauthors{J. I. Gonz\'alez Hern\'andez et al.}

\begin{document}


\title{The fast spiral-in of the companion star \\ 
to the black hole XTE J1118+480\thanks{Based on observations made 
with the Gran Telescopio Canarias (GTC), instaled in the Spanish
Observatorio del Roque de los Muchachos of the Instituto de
Astrof{\'\i}sica de Canarias, in the island of La Palma.}}


\author{Jonay I. Gonz\'alez Hern\'andez\altaffilmark{1,2}, 
Rafael Rebolo\altaffilmark{1,2,3}, and
Jorge Casares\altaffilmark{1,2}}

\altaffiltext{1}{Instituto de Astrof{\'\i }sica de Canarias (IAC), 
E-38205 La Laguna, Tenerife, Spain; jonay@iac.es, rrl@iac.es, 
jorge.casares@iac.es}
\altaffiltext{2}{Depto. Astrof{\'\i }sica, Universidad de La 
Laguna (ULL), E-38206 La Laguna, Tenerife, Spain} 
\altaffiltext{3}{Consejo Superior de Investigaciones 
Cient{\'\i}ficas, Spain} 


\begin{abstract}

We report the detection of an orbital period decay of 
$\dot P=-1.83\pm0.66$~ms~yr$^{-1}$ in the black hole X-ray 
binary \mbox{XTE J1118+480}. 
This corresponds to a period change of $-0.85\pm0.30$~$\mu $s per
orbital cycle, which is $\sim 150$~times larger
than expected from the emission of gravitational 
waves. 
These observations cannot be reproduced by conventional models 
of magnetic braking even when including 
significant mass loss from the system. 
The spiral-in of the star is either
driven by magnetic braking under extremely high magnetic fields
in the secondary star or by a
currently unknown process, which will have an impact on
the evolution and lifetime of black hole X-ray binaries.

\end{abstract}



\section{Introduction}
       
Angular momentum loss (AML) in short-period black hole X-ray
binaries are driven essentially by magnetic
braking~\citep{ver81}, gravitational
radiation~\citep{lan62,tay82}, and mass loss~\citep{rap82}, 
with possible additional contributions from jets~\citep{kin99}. 
Measurements of orbital period variations with time constrain 
the role of these processes in the evolution of such 
binaries~\citep{ver93}. 

Magnetic braking is proposed as the main mechanism 
for AML in compact binaries~\citep{rap83}. 
It produces the shrinking of the binary orbit
and maintains the donor star in contact with its Roche lobe,
therefore, sustaining stable mass transfer. 
Although this is an essential mechanism to understand compact 
binary evolution its actual prescription is currently debated
and not well established~\citep{iva06}.
 
\mbox{XTE J1118+480} is one of only 18 galactic X-ray binaries
which contain a dynamical black hole~\citep{cas07}. With an
orbital period of 4.1 hours it is the most compact black hole X-ray
binary known. 
Determination of the times at the 
inferior conjunction of the $\sim 0.2$~\Msun secondary star 
with respect to the $\sim 8$~\Msun black hole have been obtained 
at different epochs since its discovery on UT 2000 March 29 by 
the Rossi X-ray Timing Explorer~\citep{rem00}. 
Previous attempts to estimate the orbital period derivative 
failed due to the small baseline of the 
observations~\citep{joh09a}.

\section{Observations}

\begin{figure}[ht!]
\centering
\includegraphics[height=8.5cm,angle=90]{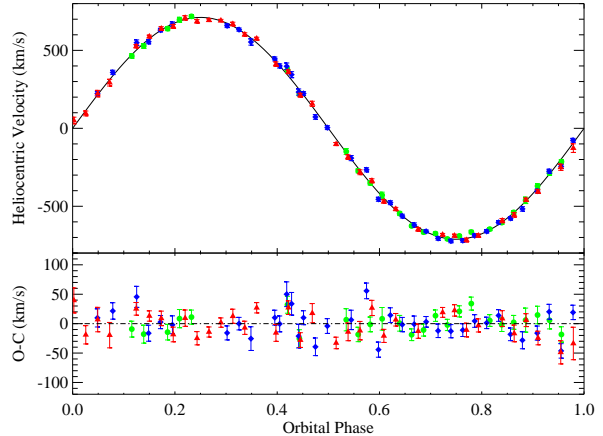}
\caption{\scriptsize{{\it Top panel}: radial velocities of the secondary star
in the black hole X-ray binary \mbox{XTE J1118+480} obtained from the
GTC/OSIRIS spectroscopic data taken on the 
three nights of UT 2011 January 7 (green filled circles), UT 2011 
February 8 (blue filled diamonds) and UT 2011 April 25 (red filled
triangles), folded on the best-fitting orbital solution. 
{\it Bottom panel}: residuals of the fit, with a rms of $\sim 20 
{\rm km}\ {\rm s}^{-1}$.}}
\label{frv}   
\end{figure}

We have conducted new spectroscopic observations of 
\mbox{XTE J1118+480} using the 10.4m Gran 
Telescopio Canarias (GTC) equipped with the OSIRIS 
spectrograph~\citep{cep00,cep03} at 
the Observatorio del Roque de los Muchachos in La Palma 
(Canary Islands, Spain). Ninety seven medium-resolution spectra
($\lambda/\delta\lambda\sim2,500$) were obtained on UT 2011 
January 7, February 8, and April 25--25, 36 and 36 spectra
in each night respectively--.
In Fig.~\ref{frv} we display the radial velocities of the secondary 
star in its orbital motion around the center-of-mass of 
the system, derived using the cross-correlation technique with a
template stellar spectrum properly broadened with a 
rotational velocity of $v\sin i =100$~\kms ~\citep{gon06,gon08}.
The radial velocity points, which spread over $\sim 3$ 
and a half months, provide a new determination of the current orbital 
period of $P_{\rm orb}=0.16993379\pm0.00000047$~d, which is smaller 
although still consistent with the orbital period measurement 
previously determined on UT 2000 December 1 ~\citep{tor04}. 
In Table~\ref{tpar} we list these orbital period measurements,
$P_{\rm orb}$, the updated dynamical masses of the black hole, 
$M_{\rm BH}$, and the secondary star, $M_2$, the mass ratio, $q$,
the rotational velocity, $v \sin i$, the orbital inclination, $i$,
the orbital semiamplitude velocity, $k_2$, the mass function, $f(M)$,
the current radius of the secondary star, $R_2$,
and orbital separation of the system components, $a_c$.

\begin{deluxetable}{lcc}
\tabletypesize{\scriptsize}
\tablecaption{Kinematical and dynamical binary parameters of 
\mbox{XTE J1118+480}.\label{tpar}}   
\tablewidth{0pt}
\tablehead{\colhead{Parameter} & \colhead{Value} & \colhead{Reference}}
\startdata
$v \sin i$         & $100^{+3}_{-11}$~\kms     & [1] \\
$i$                & $68\pm2$                  & [2] \\
$k_2$              & $708.8\pm1.4$~\kms        & [1] \\
$P_{\rm orb}$      & $0.1699339(2)$~d & [3] \\
$P_{\rm orb}$      & $0.1699338(5)$~d & [4] \\
$q=M_2/M_{\rm BH}$ & $0.027\pm0.009$           & [1] \\
$f(M)$             & $6.27\pm0.04$~\Msun       & [1] \\
$M_{\rm BH}$       & $8.30^{+0.28}_{-0.14}$~\Msun & [4,5] \\
$M_2$              & $0.22\pm0.07$~\Msun          & [4,5] \\
$R_2$              & $0.37^{+0.04}_{-0.05}$~\Rsun & [4,5] \\
$a_c$              & $2.64^{+0.03}_{-0.02}$~\Rsun & [4,5] \\
\enddata
\tablecomments{References: [1]~\citet{gon06}; [2]~\citet{gel06}; [3]
~\citet{tor04}; [4] This work; [5]~\citet{gon11}} 
\end{deluxetable}

\section{Orbital period decay}

\begin{figure}[ht!]
\centering
\includegraphics[height=8.5cm,angle=90]{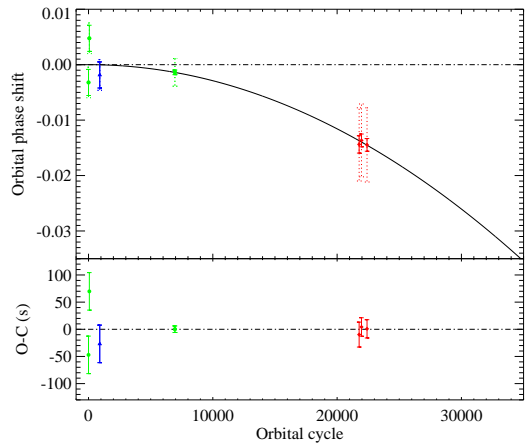}
\caption{\scriptsize{{\it Top panel}: orbital phase shift at the time of the 
inferior conjunction (orbital phase 0), $T_n$, of the secondary star in the low-mass 
black hole X-ray binary \mbox{XTE J1118+480} versus the orbital 
cycle number, $n$, folded on the best-fit parabolic fit. 
Green filled circles are spectroscopic determinations, 
the blue filled triangle is a photometric measurement and 
red diamonds are the new GTC/OSIRIS spectroscopic determinations.
Solid error bars show the uncertainties associated to 
the $T_n$ determinations, whereas the dashed error bars also 
include the uncertainties associated to the $T_0$ and $P_0$
determinations. 
{\it Bottom panel}: residuals of the fit of the $T_n$ values versus 
the cycle number $n$.}} 
\label{fph}   
\end{figure}

The spectroscopic data were used to derive three new times, 
$T_n$, of the inferior conjunction of the secondary star in 
this system (see Table~\ref{ttn}). 
Assuming a constant rate of change of the orbital
period, the time, $T_n$, of the $n$th orbital cycle can be expressed 
as $T_n=T_0+P_0 n+\frac{1}{2} P_0 \dot P n^2$, where $P_0$ is the 
orbital period at time $T_0$ of the reference cycle ($n=0$), 
$\dot P$ is the orbital period time derivative, and $n$, 
the orbital cycle number. We use the IDL routine {\scshape curvefit} 
and obtain $T_0=2451868.8921\pm0.0002$~d, 
$P_0=0.16993404\pm0.00000005$~d, and
a period derivative of $\dot P = -(5.8\pm2.1) \times 10^{-11}$ s/s 
with a reduced $\chi^2_\nu=1.7$ with $\nu=3$. A linear fit 
($\dot P =0$), and a third-order polynomial 
fit (including $\ddot P$), provide 
worse fits with $\chi^2_\nu=2.9$ and 2.3,
respectively.
In Fig.~\ref{fph} we have depicted the orbital phase
shift, defined as $\phi_n=\frac{T_n-T_0}{P_0}-n$, 
of each of the $T_n$ values as a function of the orbital cycle 
number $n$, together with the best-fit second-order solution. 
This figure shows a clear deviation from the null variation and 
that  $\dot P$ is negative.  Our result, which can be expressed as 
$\dot P=-1.83\pm0.66$~ms~yr$^{-1}$, represents the first 
determination of the orbital shrinkage in a low-mass  
black hole X-ray binary.

\section{Discussion and conclusions\label{sdisc}}

General Relativity~\citep{mis73} predicts a secular
periastron precesion period of $\sim 23$~yr for \mbox{XTE J1118+480}. 
In the case of non-zero orbital eccentricity, this 
would affect the period inferred from succesive inferior 
conjunctions ~\citep[see e.g. eq.~20 in][]{pal08}.
Using our data we set an upper-limit to the eccentricity of 
the black hole binary orbit of $e < 0.0067$ (95\% confidence
level), and conclude that relativistic periastron precesion 
could only explain up to 50\% of the measured orbital period decay.

\begin{deluxetable}{lccc}
\tabletypesize{\scriptsize}
\tablecaption{Time at inferior conjunction of the secondary star 
in \mbox{XTE J1118+480}.\label{ttn}}   
\tablewidth{0pt}
\tablehead{\colhead{$N$} & \colhead{$T_n-2450000$}\tablenotemark{a} & 
\colhead{$\delta T_n$}& \colhead{Reference}}
\startdata
0     & 1868.8916  & 0.0004 & [1] \\
66    & 1880.1086  & 0.0004 & [2] \\
904   & 2022.5122\tablenotemark{b}  & 0.0004 & [3] \\
6950  & 3049.93347 & 0.00007 & [4] \\
21772 & 5568.6936  & 0.0003 & [5] \\
21960 & 5600.6413  & 0.0002 & [5] \\
22407 & 5676.6017  & 0.0002 & [5] \\
\enddata
\tablenotetext{a}{Time at inferior conjunction, $T_n$, of the secondary 
star in the low-mass black hole X-ray binary \mbox{XTE J1118+480}, 
and uncertainties, $\delta T_n$, corresponding to a given 
orbital cycle number, $n$.}
\tablenotetext{b}{From photometric measurements.}
\tablecomments{References: [1]~\citet{wag01}; [2]~\citet{tor04}; [3]
~\citet{zur03}; [4]~\citet{gon08}; [5]~This work.} 
\end{deluxetable}

The loss of energy through emission of quadrupole 
gravitational radiation, as predicted by general relativity, provides 
an orbital period derivative for this binary system of 
$\dot P_{\rm GR} \sim -0.01$~ms~yr$^{-1}$ 
\citep[from eq.~8 in][]{tay82} 
which is far below the observed rate of variation of the orbital 
period. On the other hand, the presence of jets in 
\mbox{XTE J1118+480} has also been postulated based on X-ray and 
infrared observations~\citep[see e.g.][]{mil02}. 
However, the consequential AML due to jets even for relativistic jets \citep{kin99} 
is $\leq 1000$ times lower than the AML due to 
magnetic braking. 
The presence of a circumbinary disk is also proposed as 
possible AML mechanism in cataclysmic variables 
\citep[see e.g.][]{taa03}, and the detection of mid-IR 4.5--8~$\mu$m 
excess emission~\citep{mun06} do support its existence in 
\mbox{XTE J1118+480}. However, the estimated total mass of 
this disk is too small to affect the orbital angular momentum of 
this X-ray binary. Therefore, these effects cannot account for any
significant additional contribution to the apparently too large 
orbital period decay measured in this black hole binary.

\begin{figure}[ht!]
\centering
\includegraphics[width=8.5cm,angle=0]{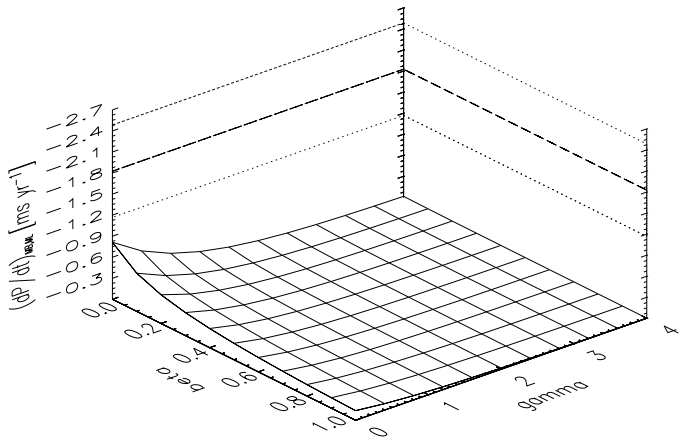}
\caption{\scriptsize{Theoretical estimate of the orbital period derivative in the 
low-mass X-ray black hole binary \mbox{XTE J1118+480} due to magnetic
braking and mass loss, 
$\dot P_{\rm MB,ML}$ given in ms/yr, derived from the second term of
eq.~28 in~\citet{joh09b}}, as a function of two parameters: 
$\beta=-\dot M_{\rm BH}/ \dot M_2$, the fraction of mass lost by 
the secondary star that is captured by the black hole; and $\gamma$, 
an index that characterizes the strength of the magnetic braking. 
The specific angular momentum, $j_w$, carried 
away by the mass lost from the system has been set to 1 which 
indicate that all the mass lost from the system is lost from the
neighborhood of the black hole 
\citep[or its accretion disk, see][]{pod02}. 
Dashed lines represent the value derived from observations and the 
dotted lines, the values at 1-$\sigma$ ver the uncertainty 
on the orbital period derivative, $\Delta\dot P$.} 
\label{fmbml}   
\end{figure}

Most conventional magnetic braking and mass loss models for 
the binary parameters of \mbox{XTE J1118+480} predict a much lower 
orbital period decay than we measure (see Fig.~\ref{fmbml}).
Our observations appear inconsistent with these 
models. Only a very restrictive, rather unplausible subset of models,
where nearly all the mass transferred by the secondary star 
is also lost by the system ($\beta=-\dot M_{\rm BH}/ \dot M_2\sim0$), may lead to 
values of the orbital period
derivative close to our observations (see Fig.~\ref{fmbml}).

Magnetic braking could be enhanced 
by anomalously high magnetic fields in the secondary star. 
 We have estimated, from eq.~5 in~\citet{jus06} and assuming that the
mass lost by wind is equal to the mass transfer rate (derived from 
eq.~9 in~\citet{kin96}), that a magnetic field at the surface of 
the companion star $B_s\ge 10-20$~kG is 
needed to explain the observations. The angular momentum loss due to
magnetic braking is determined from eq.~9
in~\citet{joh09b}.
This magnetic field is 1--2 orders of magnitude larger than typical 
magnetic field strengths in highly-rotating low-mass 
stars~\citep{pha09}. 
However, we cannot discard that the donor star in \mbox{XTE J1118+480}
descends from a magnetically peculiar Ap/Bp star that has retained 
most of its primordial magnetic field. 
In such stars very high magnetic fields of $\sim
20$~kG have been observed~\citep{elk10}. Binaries with
intermediate-mass Ap/Bp stars have indeed been postulated as the 
progenitors of compact black hole X-ray binaries~\citep{jus06}.  
In particular, magnetically coupled, irradiation driven 
stellar winds can lead to substantial loss of systemic angular 
momentum~\citep{jus06} required to form low-mass X-ray binaries
out from intermediate-mass binaries~\citep{pod02}.
The detection of CNO-processed material in \mbox{XTE J1118+480}
strongly suggests that the donor descends from an intermediate-mass
star~\citep{has02}.
Our discovery of a large period decay in XTE J1118+480 is also
consistent with this scenario although we cannot rule out the
possibility that other mechanism, not yet identified, plays a major
role in the loss of angular momentum and hence the evolution of black
hole X-ray binaries. In any case, our observations suggests a faster
evolution and shorter lifetimes than previously assumed. 
This would help to reconcile the population number of
low-mass X-ray binaries and millisecond pulsars, a long-standing
problem in galactic astronomy~\citep{pod02}. 
Follow-up spectroscopy in the coming
years may provide a determination of the second derivative of the
orbital period with strong implications on our knowledge on the
formation and evolution of black hole X-ray binaries. 

The black hole in \mbox{XTE J1118+480} was possibly formed in a
violent supernova explosion that launched the system via an
asymmetric natal kick~\citep{gua05} from its formation
region in the Galactic thin disk~\citep{gon06} to its present
location in the Galactic halo~\citep{mir01}. 
A lower limit to the age of the
system of $\geq 11$~Myr was derived from its peculiar location and 
kinematics. 
This limit has been used to set constraints on the rate at which 
black holes  can evaporate in the Anti-de Sitter (AdS) braneworld 
Randall-Sundrum gravity 
model~\citep[for details see the introduction in][]{joh09b,joh09a}
via the emission of a large number of conformal field theory (CFT) 
modes~\citep{emp03}.

An upper-limit to the asymptotic AdS curvature radius
in the extra dimensions was established at
$L\leq 80$~$\mu m$~\citep{psa07}, restricting deviations
from the gravitational inverse square law to manifest only at
distances smaller than L.

The orbital
period evolution in this system was studied before but with a less
number of measurements of $T_n$ spread in a shorter interval of time,
thus providing a period derivative consistent with zero at 1-$\sigma$
\citep{joh09a}, and an upper-limit of $L\leq 97$~$\mu m$.
 In their eq.~2, these authors assumed no angular momentum
loss due to mass lost from the system ($j_w=0$), no accretion onto 
the black hole ($\beta=0$), and the parameter $\gamma$, which 
governs the strength of magnetic braking, equal to zero 
\citep[see also Fig.~1 in][]{joh09a}. 
If we adopt the black hole and secondary masses given in 
Table~\ref{tpar}, our determination of the orbital period
decay provides a much tighter constraint on the 
asymptotic AdS curvature radius of $L\leq 35$~$\mu m$ at 2-$\sigma$. 
This limit is more restricted than the best current table-top 
experiment upper-limit of $L\leq 44$~$\mu m$ ~\citep{kap07}. 
We note here that the size of the extra dimensions has been
also recently constrained from the age of a black hole in a 
extragalactic globular cluster, placing an upper-limit as low as 
$L\leq 3$~$\mu m$ ~\citep[see][]{gne09}. 

\acknowledgments

J.I.G.H. acknowledges financial support from the Spanish Ministry 
project MICINN AYA2008-04874 and from the Spanish Ministry of
Science and Innovation (MICINN) under the 2009 Juan de la Cierva 
Programme. J.C. acknowledges MICINN grant AYA2010-18080. J.C and R.R.
are also grateful to the Consolider CSD2006-70 project of "First
Science with the GTC".
We are grateful to T. Marsh for the use of the MOLLY analysis package.
This work has made use of the IRAF facilities.

\end{document}